\documentclass[twocolumn,prl,superscriptaddress]{revtex4}

\usepackage{dcolumn}
\usepackage{amsmath}

\usepackage{graphicx}

\newcommand{\dd}{\mathrm{d}}
\newcommand{\half}{\tfrac12}
\newcommand{\set}[1]{\lbrace#1\rbrace}
\newcommand{\etal}{{\it{}et~al.}}
\newcommand{\mat}{\mathbf}
\renewcommand{\vec}{\mathbf}

\newcommand\pin{p_\textrm{in}}
\newcommand\pout{p_\textrm{out}}
\newcommand\cin{c_\textrm{in}}
\newcommand\cout{c_\textrm{out}}

\begin{document}

\title{Generalized communities in networks}
\author{M. E. J. Newman}
\affiliation{Department of Physics and Center for the Study of Complex
  Systems, University of Michigan, Ann Arbor, MI 48109, U.S.A.}
\affiliation{Santa Fe Institute, 1399 Hyde Park Road, Santa Fe, NM 87501,
  U.S.A.}
\author{Tiago P. Peixoto}
\affiliation{Institut f\"ur Theoretische Physik, Universit\"at Bremen,
  Hochschulring 18, D-28359 Bremen, Germany}

\begin{abstract}
  A substantial volume of research has been devoted to studies of community
  structure in networks, but communities are not the only possible form of
  large-scale network structure.  Here we describe a broad extension of
  community structure that encompasses traditional communities but includes
  a wide range of generalized structural patterns as well.  We describe a
  principled method for detecting this generalized structure in empirical
  network data and demonstrate with real-world examples how it can be used
  to learn new things about the shape and meaning of networks.
\end{abstract}

\maketitle

The detection and analysis of large-scale structure in networks has been
the subject of a vigorous research effort in recent years, in part because
of the highly successful application of ideas drawn from statistical
physics~\cite{Boccaletti06,Newman10}.  Particular energy has been devoted
to the study of community structure, meaning the division of networks into
densely connected subgroups, a common and revealing feature, especially in
social and biological networks~\cite{Fortunato10}.  Community structure is,
however, only one of many possibilities where real-world networks are
concerned.  In this paper, we describe a broad generalization of community
structure that encompasses not only traditional communities but also
overlapping or fuzzy communities, ranking or stratified structure,
geometric networks, and a range of other structural types, yet is easily
and flexibly detected using a fast, mathematically principled procedure
which we describe.  We give demonstrative applications of our approach to
both computer-generated test networks and real-world examples.

Community structure can be thought of as a division of the nodes of a
network into disjoint groups such that the probability of an edge is higher
between nodes in the same group than between nodes in different groups.
For instance, one can generate artificial networks with community structure
using the stochastic block model, a mathematical model that follows exactly
this principle.  In the stochastic block model the nodes of a network are
divided into $k$ groups, with a node being assigned to group~$r$ with some
probability~$\gamma_r$ for $r=1\ldots k$, and then edges are placed between
node pairs independently with probabilities $p_{rs}$ where~$r$ and~$s$ are
the groups the nodes fall in.  If the diagonal probabilities~$p_{rr}$ are
larger than the off-diagonal ones, we get traditional community structure.

Alternatively, however, one can also look at the stochastic block model
another way: imagine that we assign each node a random node parameter~$x$
between zero and one and edges are placed between node pairs with a
probability~$\omega(x,y)$ that is a function of the node parameters $x$ and
$y$ of the pair.  If $\omega(x,y)$ is piecewise constant with $k^2$
rectangular regions of size $\gamma_r\gamma_s$ and value~$p_{rs}$, then
this model is precisely equivalent to the traditional block model.  But
this prompts us to ask what is so special about piecewise constant
functions?  It is certainly possible that some networks might contain
structure that is better captured by functions~$\omega(x,y)$ of other
forms.  Why not let $\omega(x,y)$ take a more general functional form,
thereby creating a generalized type of community structure that includes
the traditional type as a subset but can also capture other structures as
well?  This is the fundamental idea behind the generalized structures of
this paper: edge probabilities are arbitrary functions of continuous node
parameters.

The idea is related to two threads of work in the previous literature.
One, in sociology and statistics, concerns ``latent space'' models, in
which nodes in a network are located somewhere in a Euclidean space and are
more likely to be connected if they are spatially close than if they are
far apart~\cite{HRH02}.  A specific functional form is typically assumed
for the connection probability and the model is fitted to data using Monte
Carlo methods.  The other thread, in the mathematics literature, concerns
so-called ``graphon'' models and does not deal with the analysis of
empirical data but with the mathematical properties of models, showing in
particular that models of a kind similar to that described here are
powerful enough to capture the properties of any theoretical ensemble of
networks in the limit of large size, at least in the case where the
networks are dense~\cite{Lovasz12,BCCZ14}.

In this paper, we define a specific model of generalized community
structure and a method for fitting it to empirical data using Bayesian
inference.  The fit places each node of the network in the ``latent space''
of the node parameters~$x$ and, simultaneously, gives us an estimate of the
probability function~$\omega(x,y)$.  Between them, these two outputs tell
us a great deal about the structure a network possesses and the role each
node plays within that structure.  The method is computationally efficient,
allowing for its application to large networks, and provides significantly
more insight than the traditional community division into discrete groups,
or even recent generalizations to overlapping groups~\cite{PDFV05,ABFX08}.

We begin by defining a model that generates networks with the generalized
community structure we are interested in.  The model follows the lines
sketched above, but with some crucial differences.  We take $n$ nodes and
for each node~$u$ we generate a node parameter~$x_u$ uniformly at random in
the interval~$[0,1]$.  Then between each pair of nodes~$u,v$ we place an
undirected edge with probability
\begin{equation}
p_{uv} = {d_u d_v\over 2m}\,\omega(x_u,x_v),
\label{eq:puv}
\end{equation}
where $d_u,d_v$ are the degrees of the nodes, $m=\half \sum_u d_u$ is the
total number of edges in the network, and $\omega(x,y)$ is a function of
our choosing, which we will call the \textit{edge function}.  Note that
$\omega(x,y)$ must be symmetric with respect to its arguments for an
undirected network such as this.

The inclusion of the degrees is necessary to accommodate the fitting of
networks with broad degree distributions (which includes essentially all
real-world networks~\cite{BA99b,ASBS00}).  Without it, the model
effectively assumes a Poisson degree distribution, which is a poor fit to
most networks and can cause the calculation to fail~\cite{KN11a}.  The
factor $d_u d_v/2m$ is the probability of an edge between nodes with
degrees~$d_u,d_v$ if edges are placed at random~\cite{Newman10}.  Hence
$\omega(x_u,x_v)$ parametrizes the variation of the probability relative to
this baseline level and is typically of order~1, making $p_{uv}$ small in
the limit where~$m$ becomes large.

Given the model, we fit it to empirical network data using the method of
maximum likelihood.  The probability or
likelihood~$P(\mat{A},\vec{x}|\omega)$ that we generate a particular set of
node parameters~$\vec{x}=\set{x_u}$ and a particular network structure
described by the adjacency matrix~$\mat{A}=\set{a_{uv}}$ is
\begin{equation}
P(\mat{A},\vec{x}|\omega) = \prod_{u<v} p_{uv}^{a_{uv}} (1-p_{uv})^{1-a_{uv}}.
\label{eq:likelihood}
\end{equation}
To find the value of the edge function~$\omega(x,y)$ that best fits an
observed network we want to maximize the marginal likelihood
\begin{equation}
P(\mat{A}|\omega) = \int P(\mat{A},\vec{x}|\omega) \>\dd^n\vec{x},
\label{eq:marginal}
\end{equation}
or equivalently its logarithm, whose maximum falls in the same place.
Direct maximization leads to a set of implicit equations that are hard to
solve, even numerically, so instead we employ the following trick.

For any positive-definite function~$f(x)$, Jensen's inequality says that
\begin{equation}
\log \int f(x) \>\dd x \ge \int q(x) \log {f(x)\over q(x)} \>\dd x,
\label{eq:jensen}
\end{equation}
where $q(x)$ is any probability distribution over~$x$ such that $\int q(x)
\>\dd x = 1$.  Applying~\eqref{eq:jensen} to the log of the marginal
likelihood, Eq.~\eqref{eq:marginal}, we get
\begin{equation}
\log \int P(\mat{A},\vec{x}|\omega) \>\dd^n\vec{x}
  \ge \int q(\vec{x}) \log {P(\mat{A},\vec{x}|\omega)\over q(\vec{x})}
   \>\dd^n\vec{x},
\label{eq:maximization}
\end{equation}
where $q(\vec{x})$ is any probability distribution over~$\vec{x}$.  It is
straightforward to verify that the exact equality is recovered, and hence
the right-hand side maximized, when
\begin{equation}
q(\vec{x}) = {P(\mat{A},\vec{x}|\omega)\over
              \int P(\mat{A},\vec{x}|\omega) \>\dd^n\vec{x}}.
\label{eq:qx}
\end{equation}
Further maximization with respect to~$\omega$ then gives us the maximum of
the marginal likelihood, which is the result we are looking for.  Put
another way, a double maximization of the right-hand side
of~\eqref{eq:maximization} with respect to both~$q(\vec{x})$ and~$\omega$
will achieve the desired result.  And this double maximization can be
conveniently achieved by alternately maximizing with respect
to~$q(\vec{x})$ using~\eqref{eq:qx} and with respect to~$\omega$ by
differentiating.

This method, which is a standard one in statistics and machine learning, is
called an expectation--maximization or EM algorithm~\cite{MK08}.  It
involves simply iterating these two operations from (for instance) a random
initial condition until convergence.  The converged value of the
probability density~$q(\vec{x})$ has a nice physical interpretation.
Combining Eqs.~\eqref{eq:marginal} and~\eqref{eq:qx}, we have $q(\vec{x}) =
P(\mat{A},\vec{x}|\omega)/P(\mat{A}|\omega) = P(\vec{x}|\mat{A},\omega)$.
In other words, $q(\vec{x})$~is the posterior probability distribution on
the node parameters~$\vec{x}$ given the observed network and the edge
function~$\omega(x,y)$.  It tells us the probability of any given
assignment~$\vec{x}$ of parameters to nodes.  It is this quantity that will
in fact be our primary object of interest here.

Substituting from Eqs.~\eqref{eq:puv} and~\eqref{eq:likelihood}
into~\eqref{eq:maximization}, keeping terms to leading order in small
quantities and dropping overall constants, we can write the quantity to be
maximized as
\begin{equation}
\iint_0^1 \sum_{uv} q_{uv}(x,y) \biggl[ a_{uv} \log \omega(x,y)
  - {d_u d_v \omega(x,y)\over2m} \biggr] \>\dd x\>\dd y,
\label{eq:objective}
\end{equation}
where $q_{uv}(x,y) = \int q(\vec{x}) \delta(x_u-x) \delta(x_v-y)
\>\dd^n\vec{x}$ is the posterior marginal probability that nodes $u,v$ have
node parameters~$x,y$ respectively.  The obvious next step is to
maximize~\eqref{eq:objective} by functional differentiation with respect
to~$\omega(x,y)$, but there is a problem.  If we allow $\omega$ to take any
form at all then it has an infinite number of degrees of freedom, which
guarantees overfitting of the data.  Put another way, physical intuition
suggests that $\omega(x,y)$ should be smooth in some sense, and we need a
way to impose that smoothness as a constraint on the optimization.  There
are a number of ways we could achieve this, but a common one is to express
the function in terms of a finite set of basis functions.  For nonnegative
functions such as $\omega$ a convenient basis is the Bernstein polynomials
of degree~$N$:
\begin{equation}
B_k(x) = {N\choose k} x^k (1-x)^{N-k},\qquad k=0\ldots N.
\end{equation}
The Bernstein polynomials form a complete basis for polynomials of
degree~$N$ and are nonnegative in $[0,1]$, so a linear combination
$\sum_{k=0}^N c_k B_k(x)$ is also nonnegative provided $c_k\ge0$ for
all~$k$.  Our edge function~$\omega(x,y)$ is a function of two variables,
so we will write it as a double expansion in Bernstein polynomials
\begin{equation}
\omega(x,y) = \sum_{j,k=0}^N c_{jk} B_j(x) B_k(y),
\label{eq:bernstein}
\end{equation}
which again is nonnegative for $c_{jk}\ge0$.  Bernstein polynomials have
excellent stability properties under fluctuations of the values of the
expansion coefficients, which makes them ideal for statistical applications
such as ours.  Note that since $\omega(x,y)$ is symmetric with respect to
its arguments we must have $c_{jk}=c_{kj}$.

If $\omega(x,y)$ is constrained to take this form, then instead of the
unconstrained maximization of~\eqref{eq:objective} we now want to maximize
with respect to the coefficients~$c_{jk}$.  To do this, we substitute
from~\eqref{eq:bernstein} into~\eqref{eq:objective} and apply Jensen's
inequality again, this time in its summation form $\log \sum_i f_i \ge
\sum_i Q_i \log f_i/Q_i$.  Then, by the same argument as previously, we
find that the optimal coefficient values are given by the double
maximization with respect to $c_{jk}$ and~$Q_{jk}(x,y)$ of
\begin{align}
& \iint_0^1 \mu(x,y) \sum_{jk} Q_{jk}(x,y)
  \log {c_{jk} B_j(x) B_k(y)\over Q_{jk}(x,y)} \>\dd x\>\dd y \nonumber\\
&\qquad{} - \iint_0^1 \nu(x) \nu(y) \sum_{jk} c_{jk} B_j(x) B_k(y)
  \>\dd x \>\dd y,
\end{align}
where
\begin{equation}
\mu(x,y) = {1\over2m} \sum_{uv} a_{uv} q_{uv}(x,y),\quad
\nu(x) = {1\over2m} \sum_u d_u q_u(x),
\label{eq:munu}
\end{equation}
and $q_u(x) = n^{-1} \sum_v \int q_{uv}(x,y) \>\dd y$ is the marginal
probability that node~$u$ has node parameter~$x$.  The maximization with
respect to~$Q_{jk}(x,y)$ is achieved by setting
\begin{equation}
Q_{jk}(x,y) = {c_{jk} B_j(x) B_k(y)\over\sum_{jk} c_{jk} B_j(x) B_k(y)},
\label{eq:Q}
\end{equation}
and the maximization with respect to~$c_{jk}$ is achieved by
differentiating, which gives
\begin{equation}
c_{jk} = {\iint \mu(x,y) Q_{jk}(x,y) \>\dd x\>\dd y\over
          \int \nu(x) B_j(x) \>\dd x \int \nu(y) B_k(y) \>\dd y}.
\label{eq:cjkiter}
\end{equation}
Since all quantities on the right of this equation are nonnegative,
$c_{jk}\ge0$ for all $j,k$ and hence $\omega(x,y)\ge0$, as required.

The calculation of the optimal values of the~$c_{jk}$ is a matter of
iterating Eqs.~\eqref{eq:Q} and~\eqref{eq:cjkiter} to convergence, starting
from the best current estimate of the coefficients.  Note that the
quantities $\mu$ and $\nu$ need be calculated only once each time around
the EM algorithm, and both can be calculated in time linear in the size of
the network in the common case of a sparse network with $m\propto n$.  The
integrals in Eq.~\eqref{eq:cjkiter} we perform numerically, using standard
Gauss--Legendre quadrature.

This, in principle, describes a complete algorithm for fitting the model to
observed network data, but in practice the procedure is cumbersome because
of the denominator of Eq.~\eqref{eq:qx}, which involves an $n$-dimensional
integral, where $n$ is the number of nodes in the network, which is
typically large.  The traditional solution to this problem is to subsample
the distribution~$q(\vec{x})$ approximately using Monte Carlo importance
sampling.  Here, however, we use a different approach proposed recently by
Decelle~\etal~\cite{DKMZ11a}, which employs belief propagation and returns
good results while being markedly faster than Monte Carlo.  The method
focuses on a function~$\eta_{u\to v}(x)$, called the belief, which
represents the probability that node~$u$ has node parameter~$x$ if node~$v$
is removed from the network.  The removal of node~$v$ allows us to write a
self-consistent set of equations whose solution gives us the beliefs.  The
equations are a straightforward generalization to the present model of
those given by Decelle~\etal:
\begin{align}
\eta_{u\to v}(x) &= {1\over Z_{u\to v}} \exp \biggl( - \sum_w d_u d_w
  \int_0^1 q_w(y) \omega(x,y) \>\dd y \biggr) \nonumber\\
 &\qquad{}\times \prod_{\substack{w(\ne v)\\a_{uw}=1}}
  \int_0^1 \eta_{w\to u}(y) \omega(x,y) \>\dd y,
\label{eq:bp1}
\end{align}
where $q_w(y)$ is again the marginal posterior probability for node~$w$ to
have node parameter~$y$ and as before we have dropped terms beyond leading
order in small quantities.  The quantity $Z_{u\to v}$ is a normalizing
constant which ensures that the beliefs integrate to unity:
\begin{align}
Z_{u\to v} &= \int_0^1 \exp\biggl( - \sum_w d_u d_w \int_0^1
  q_w(y) \omega(x,y) \>\dd y \biggr) \nonumber\\
 &\qquad{}\times \prod_{\substack{w(\ne v)\\a_{uw}=1}}
  \int_0^1 \eta_{w\to u}(y) \omega(x,y) \>\dd y \>\dd x.
\label{eq:bp2}
\end{align}

The belief propagation method consists of the iteration of these equations
to convergence starting from a suitable initial condition (normally the
current best estimate of the beliefs).  The equations are exact on networks
that take the form of trees, or on locally tree-like networks in the limit
of large network size (where local neighborhoods of arbitrary size are
trees).  On other networks, they are approximate only, but in practice give
excellent results.

Once we have the values of the beliefs, the crucial two-node marginal
probability $q_{uv}(x,y)$ is given by
\begin{equation}
q_{uv}(x,y) = {\eta_{u\to v}(x) \eta_{v\to u}(y) \omega(x,y)\over
               \iint_0^1 \eta_{u\to v}(x) \eta_{v\to u}(y) \omega(x,y)
               \>\dd x\>\dd y}.
\label{eq:quvxy}
\end{equation}
Armed with these quantities for every node pair connected by an edge, we
can evaluate $\mu(x,y)$ and $\nu(x)$ from Eq.~\eqref{eq:munu} then iterate
Eqs.~\eqref{eq:Q} and~\eqref{eq:cjkiter} to compute new values of the
parameters~$c_{jk}$, and repeat.

We give three example applications of our methods, one to a
computer-generated benchmark network and the others to real-world networks
displaying nontrivial latent-space structure that is readily uncovered by
our algorithm.

\begin{figure}
\includegraphics[width=\columnwidth]{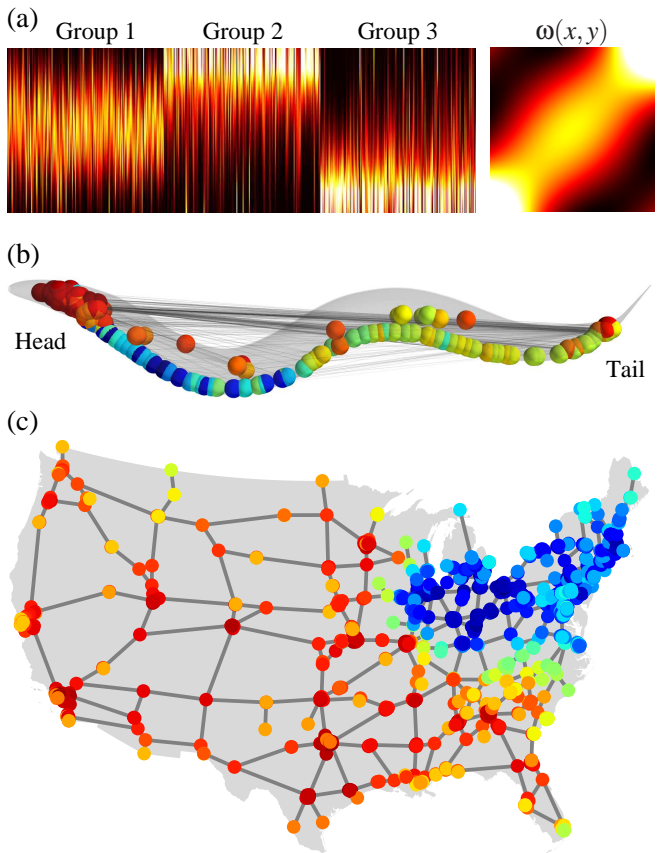}
\caption{(a)~Left: density plot of the posterior marginal probability
  densities~$q_u(x)$ that node~$u$ has node parameter~$x$ for an
  application of our algorithm to a 600-node stochastic block model with
  three groups.  Colors indicate the probabilities and there are 600
  columns, one for each node.  Right: density plot of the edge
  function~$\omega(x,y)$.  (b)~The neural network of the worm \textit{C.\
    elegans}, drawn in real space, as it falls within the body of the worm.
  Colors represent the average values of the node parameters~$x_u$ inferred
  for each neuron by our algorithm.  (c)~Network representation of the
  interstate highways of the contiguous United States.  Again, node colors
  represent the average node parameters~$x_u$.}
\label{fig:fig1}
\end{figure}

For our first example, we use a computer-generated test network created
using the standard stochastic block model, with $n=600$ nodes divided into
three equally sized groups of 200 nodes each, with probabilities
$\pin=\cin/n$ and $\pout=\cout/n$ for edges between nodes in the same and
different groups respectively and $\cin=15$, $\cout=3$.
Figure~\ref{fig:fig1}a shows a density plot of the marginal probability
distributions~$q_u(x)$ on the node parameters calculated by our algorithm
using a degree-4 (quartic) polynomial representation of the edge
function~$\omega$.  (We also used quartic representations for the other
examples below.)  The plot consists of 600 columns, one for each node,
color coded to show the value of $q_u(x)$ for the corresponding node.  As
the plot shows, the algorithm has found the three known groups in the
network, placing them at three widely spaced points in the latent space of
the node parameters.  (In this case, the first group is placed in the
middle, the second at the top, and the third at the bottom, but all orders
are equivalent.)  We also show a plot of the inferred edge
function~$\omega(x,y)$, which in this case has a heavy band along the
diagonal, indicating ``assortative'' structure, in which nodes are
primarily connected to others in the same group.

Our second example is a real-world network, the neural network of the
nematode (roundworm) \textit{C.\ elegans}, which has been mapped in its
entirety using electron microscopy~\cite{WSTB86,Szigeti14} and contains a
total of 299 neurons.  The worm has a long tubular body, with neurons
arranged not just in its head but along its entire length.  Neurons tend to
be connected to others near to them, so we expect spatial position to play
the role of a latent variable and our algorithm should be able to infer the
positions of neurons by examining the structure of network.
Figure~\ref{fig:fig1}b shows that indeed this is the case.  The figure
shows the network as it appears within the body of the worm, with nodes
colored according to the mean values of the node parameters found by the
algorithm, and we can see a strong correlation between node color and
position.  The largest number of nodes is concentrated in the head, mostly
colored red in the figure; others along the body appear in blue and green.
If we did not know the physical positions of the nodes in this case, or if
we did not know the correlation between position and network structure, we
could discover it using this analysis.

Our third example, shown in Fig.~\ref{fig:fig1}c, is an analysis of the
network of interstate highways in the contiguous United States.  This
network is embedded in geometric space, the surface of the Earth.  Again we
would expect our algorithm to find this embedding and indeed it does.  The
colors of the nodes represent the mean values of their node parameters and
there is a clear correspondence between node color and position, with the
inferred node parameters being lowest in the north-east of the country and
increasing in all directions from there.  (Note that even though the
portion of the network colored in red and orange appears much larger than
the rest, it is in fact about the same size in terms of number of nodes
because of the higher density of nodes in the north-east.)  The true
underlying space in this case has two dimensions, where our model has only
one, and this suggests a potential generalization to latent spaces with two
(or more) dimensions.  It turns out that such a generalization is possible
and straightforward, but we leave the developments for future work.

To summarize, we have in this paper described a generalized form of
community structure in networks in which, instead of placing network nodes
in discrete groups, we place them at positions in a continuous space and
edge probabilities depend in a general manner on those positions.  We have
given a computationally efficient algorithm for inferring such structure
from empirical network data, based on a combination of an EM algorithm and
belief propagation, and find that it successfully uncovers nontrivial
structural information about both artificial and real networks in example
applications.

The authors thank Cris Moore for useful discussions.  This research was
funded in part by the US National Science Foundation under grants
DMS--1107796 and DMS--1407207 and by the University of Bremen under funding
program~ZF04.

\end{document}